# Few-Layer C$_2$N: A Promising Metal-free Photocatalyst for Water Splitting

*Ruiqi Zhang,[a] and Jinlong Yang[a, b, *]*

[a] Hefei National Laboratory for Physical Sciences at the Microscale, University of Science and Technology of China, Hefei, Anhui 230026, China

[b] Synergetic Innovation Center of Quantum Information & Quantum Physics, University of Science and Technology of China, Hefei, Anhui 230026, China

**ABSTRACT**

Successful synthesis of the nitrogenated holey two-dimensional structures C$_2$N (*Nat. Commun.* **2015**, *6*, 6486) using simply wet-chemical reaction offer a cost-effective way to generate other 2D materials with novel optical and electronic properties. On basis of the density functional theory calculations, we investigate electronic properties of monolayer and multilayer C$_2$N. We find that few-layer C$_2$N have a direct bandgap and the direct bandgap of the system can vary from 2.47 eV for monolayer to 1.84 eV for a five-layer. Besides, for the few-layer C$_2$N, appropriate band gap, band edge alignments, and strong visible-light absorption demonstrate it may be a potential metal-free visible-light driven photocatalyst for water splitting.

## 1. INTRODUCTION

As increasing global energy demands and increasingly serious environment problems caused by using fossil fuels, seeking sustainable and clean alternative energy sources is urgently needed to be solved. With no reliance on fossil fuels and no carbon dioxide emission, the production of hydrogen from water using a photocatalyst and solar energy has attracted considerable attention as a potential means of renewable energy production.

Since photocatalytic splitting of water on n-type TiO2 electrodes was discovered by Fujishima in 1972,[1] varieties of semiconductors based photocatalysts for water splitting have been investigated. Unfortunately, despite intense efforts during the past few years, most of photocatalysts are discovered in metal oxides, sulfides and nitrides with $d^0$ or $d^{10}$ transition metal cations.[2–10] Most of them still face several challenging issues, such as the inability to utilize visible light, suffering from low quantum efficiency, poor stability and the cost of expensive.[11,12] So developing high efficiency and stable photocatalytic materials for water splitting has been on going.[13]

Besides the conventional photocatalytic materials containing metal, Wang et al. have reported graphitic carbon nitrides (g-$C_3N_4$), with a band gap of 2.7 eV to absorb blue light, as a metal-free photocatalyst for visible-light driven water splitting, which opens up an new way to searching for more metal-free photocatalysts.[14] However, this material is reported to show very poor quantum efficiency, due to the fast recombination of photoinduced electron-hole pairs. Then, to overcome the problem, many groups attempted to optimize the catalytic properties of g-$C_3N_4$ though texture modification, elemental doping and making g-C3N4-based composites[15–17], which were proved not to be very successful.

Besides, many and new metal-free are designed by theoretical method, like F-BNBN-H sheet,[18] semihydrogenated BN Sheet,[19] B–C–N hybrid porous sheet[20] and porous g-CN nanosheet[21] and so on, the difficult for them is being fabricated in experiment. On the other hand, many new method to improve catalytic properties of metal-free photocatalyst are reported in experiment.[22] All in all, searching for high efficiency and stable metal photocatalytic materials for water splitting is never stop.

Recently, a thinnest layered 2D crystal named $C_2N$, with uniform holes and nitrogen atoms, can be simply synthesized via a bottom-up wet-chemical reaction. Furthermore, a FET based on layer $C_2N$ with an high on/off ratio of $10^7$ was fabricated and $C_2N$ possess a optical bandgap of ~1.96 eV.[23] These results may make layer $C_2N$ a very promising candidate material for future applications in nanoscale electronics and optoelectronics.

Here, we perform first-principles to investigate monolayer and multilayer $C_2N$ as a potential metal-free photocatalyst for water splitting. To calculate the band structures more accurately, we have employed hybrid density functionals. First, we found the monolayer $C_2N$ with a direct band gap of ~2.47 eV at the Γ points based on our calculations. Then, we explored the electronic structures of the most stable multilayer $C_2N$, the result of which show that the multilayer $C_2N$ also keep a direct band gap. The band alignments of monolayer and multilayer $C_2N$ with respect to the water redox levels show that the layered $C_2N$ are satisfy for the overall water splitting. What is more, we explored that multilayer $C_2N$ has good visible-light adsorption. All in all, our results show that bilayer $C_2N$ may be a good metal-free photocatalyst candidate for water splitting.

## 2. METHODS

In this study, our first-principles calculations are based on the density functional theory (DFT) implemented in the VASP package.[24,25] The generalized gradient approximation of Perdew, Burke, and Ernzerhof (GGA-PBE)[26] and projector augmented wave (PAW) potentials are used. In all computations, the kinetic energy cutoff are set to be 520 eV in the plane-wave expansion. The first Brillouin zone is sampled with Monkhorst-Pack grid of $5\times5\times1$ for structure optimization. A large value ~15 Å of the vacuum region is used to avoid interaction between two adjacent periodic images. All the geometry structures are fully relaxed until energy and forces are converged to $10^{-5}$ eV and 0.01 eV/Å, respectively. Effect of van der Waals (vdW) interaction is accounted for by using empirical correction method proposed by Grimme (DFT-D2),[27] which is a good description of long-range vdW interactions.[28–30] As a benchmark, DFT-D2 calculations give an interlayer distance of 3.25 Å and a binding energy of -25 meV per carbon atom for bilayer graphene, consistent with previous experimental measurements and theoretical studies.[31,32] Moreover, we introduced the parameter-free Heyd−Scuseria−Ernzerhof (HSE06) hybrid functional[33,34] to precisely evaluate the band gap of layer $C_2N$.

**RESULTS AND DISCUSSION**

At first, we explored the geometric properties and electronic structure of monolayer $C_2N$. The optimized atomistic ball-stick models of unit monolayer $C_2N$ is illustrated in Fig. 1(a). Their plane structures are fully relaxed according to the force and stress calculated by DFT within the PBE functional. The equivalent lattice parameters of monolayer $C_2N$ is 8.328 Å, generating the in-plane covalent bond lengths as C-C of 1.429/1.470 Å and C-N of 1.336 Å. The band gap of single layer $C_2N$ through the HSE06 functional is a semiconductor with a direct band gap of 2.47 eV at the Γ points, which is depicted in Fig. 1(b). In the view of the band structure, unlike g-$C_3N_4$ with many localized band edges in band structure which is considered to be the reason for its low quantum efficiency,[21,35] it is clear for monolayer $C_2N$ that both the valence and conduction bands are well dispersed and no localized states are present to act as recombination centers for the photogenerated electron hole pairs. Hence the lifetime of the photogenerated charge carriers in monolayer $C_2N$ may be more longer, which can lead to better photocatalytic efficiency. At the same time, we plot the isosurfaces of band decomposed charge density corresponding to the valence band maximum (VBM) and conduction band minimum (CBM), which is shown in Fig. 1(c) and (d), respectively. Both the distribution of VBM and CBM are mainly originates from the nitrogen $P_z$ states and the C-C antibonding π states.

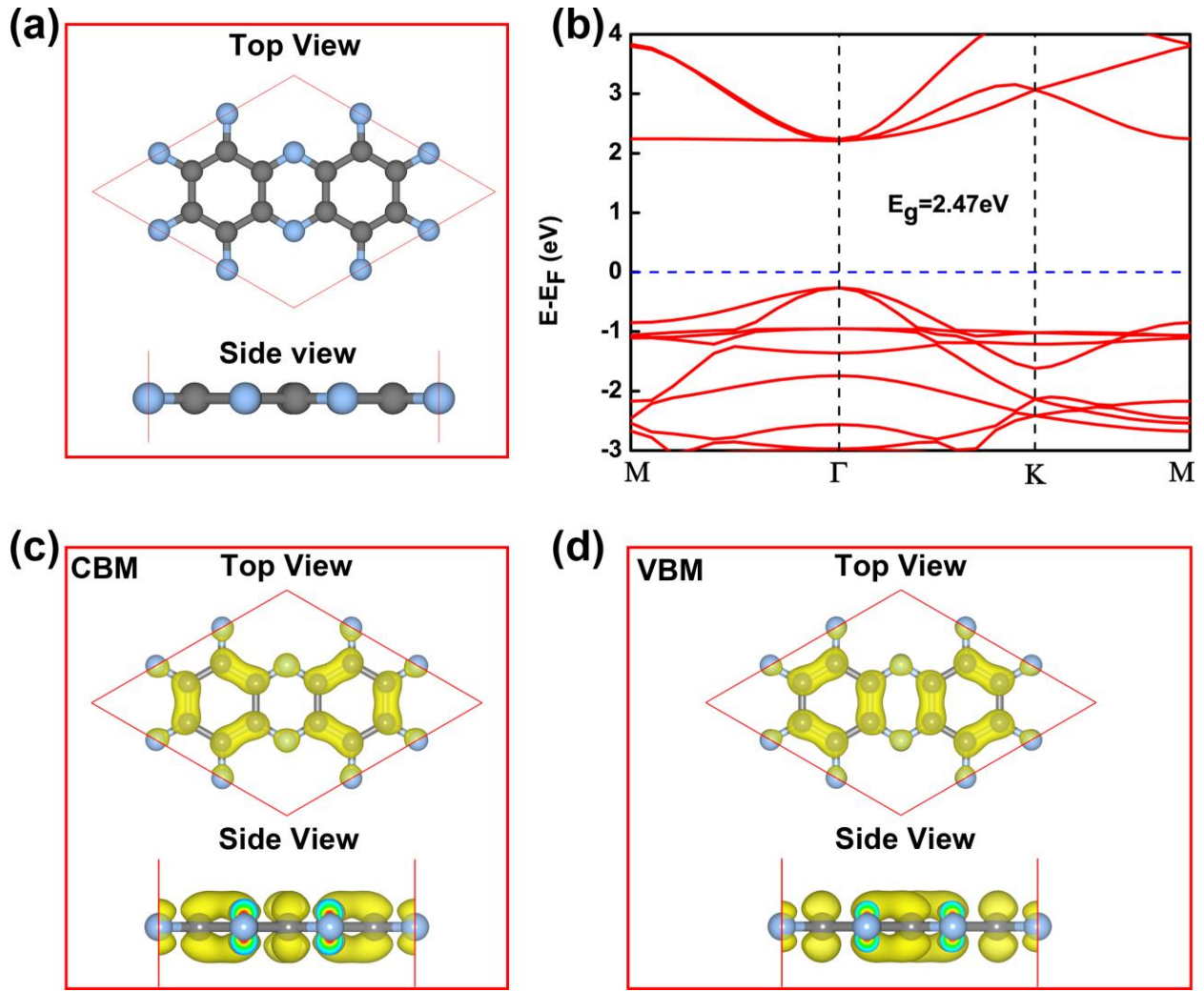

Figure 1. (a) Top view and side view of the atomic structure of monolayer C2N; (b) Band structures of monolayer C2N calculated by HSE06. (c,d) are the charge density corresponding to the CBM and VBM for monolayer, respectively. The isovalue is 0.008e/Bohr$^3$. The grey and blue ball represent C atoms and N atoms, respectively. Γ (0.0, 0.0, 0.0), M(0.5, 0.5, 0.0), and K (1/3, 1/3, 0.0) refer to special points in the first Brillouin zone.

To explore effect of interlayer coupling on the electronic structure of multilayer $C_2N$, we carried out a systematic investigation of electronic structure of bilayer, trilayer, fourlayer and fivelayer $C_2N$ using the HSE06 functionals. We only considered the most stable few-layer $C_2N$ according to our another work. The HSE06 electronic band structures of bilayer, trilayer, fourlayer and fivelayer $C_2N$ are shown in Fig. 2(a)-(d). It is clear that the direct-gap feature is well kept in few-layer $C_2N$ and both VBM and CBM are located at the $\Gamma$ point and the band gap of this system can be tuned from 2.47 eV for a monolayer to 1.84 eV for a five-layer sample. All of the bandgap are within the range of visible light, indicating the potential of bilayer $C_2N$ for photocatalyst.

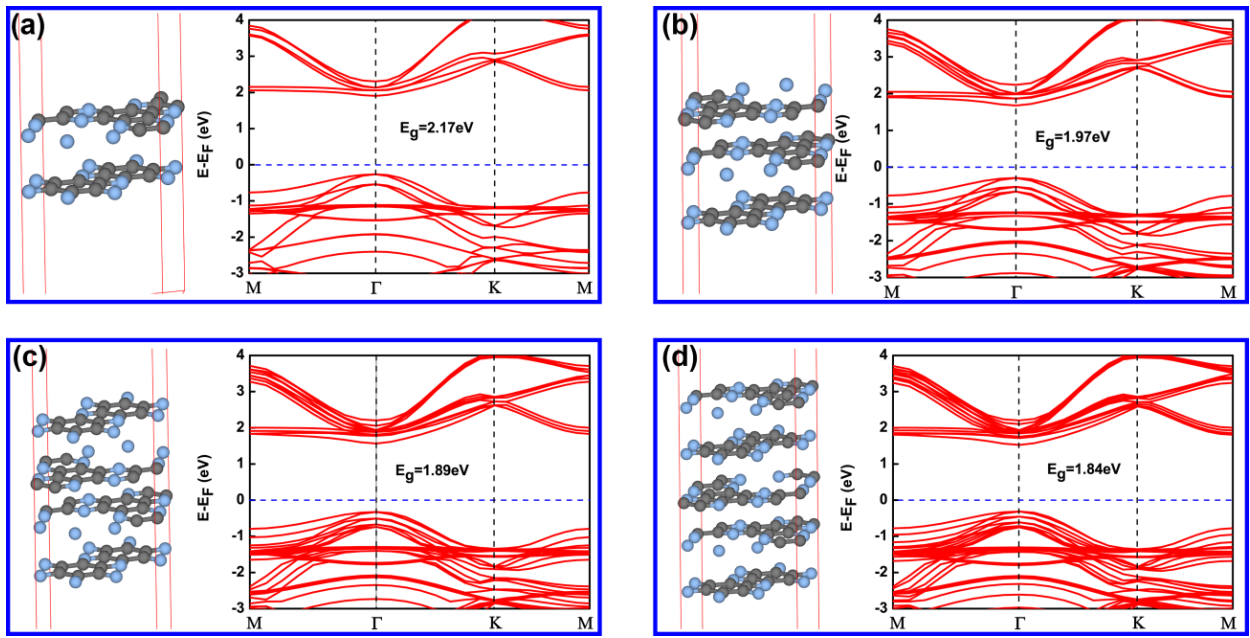

Figure 2. (a)-(d) The atomic structures and the electronic band structures of the bi-, tri-, four-, and five- layer $C_2N$, respectively. Band structures of multilayer $C_2N$ calculated by HSE06. $\Gamma$ (0.0, 0.0, 0.0), M(0.5, 0.5, 0.0), and K (1/3, 1/3, 0.0) refer to special points in the first Brillouin zone.

Another predominate factor for photocatalysis is that their band edge position need fit water reduction and oxidation potential. The reduction/oxidation ability could be evaluated by aligning the CBM and VBM with respect to the water redox potential levels. The band edge alignments of monolayer and multilayer $C_2N$ are plotted in Fig. 3(a). The standard water reduction and oxidation potential levels were marked for reference. We can note that the CBM potentials of monolayer and multilayer $C_2N$ are higher than the reduction potential of hydrogen and all of their VBM potential are lower than the oxidation potential of $O_2/H_2O$. Briefly, from thermodynamic aspect, the monolayer and multilayer $C_2N$ are good candidate as photocatalyst for water splitting.

As one of the preconditions of a high-efficiency visible light-sensitive photocatalyst, it is important for them that absorbing visible light to enhance their efficiency. Then we plot their optical absorption spectra Fig. 3(b). Clearly, multilayer $C_2N$ show substantial adsorption both in the visible light and UV light range, while for monolayer $C_2N$, its adsorption in the visible light is weak. What is more, owing to the band gap decreasing as increasing the number of layer, the absorption of few-layer $C_2N$ in the visible range getting more efficient.

As discussed above, possessing strong visible-light absorption, appropriate band-edge positions, and direct band gap, few-layer $C_2N$ may be a good candidate for promising metal-free photocatalysts. At the same time, for g-$C_3N_4$ much interest has been devoted to its based composite photocatalysts and their application for hydrogen generation under visible light.[16] Based our results, like g-$C_3N_4$, we also hope many $C_2N$ based composite photocatalysts can be studied in various aspects to further advance the utilization of $C_2N$ for visible-light photocatalytic hydrogen generation.

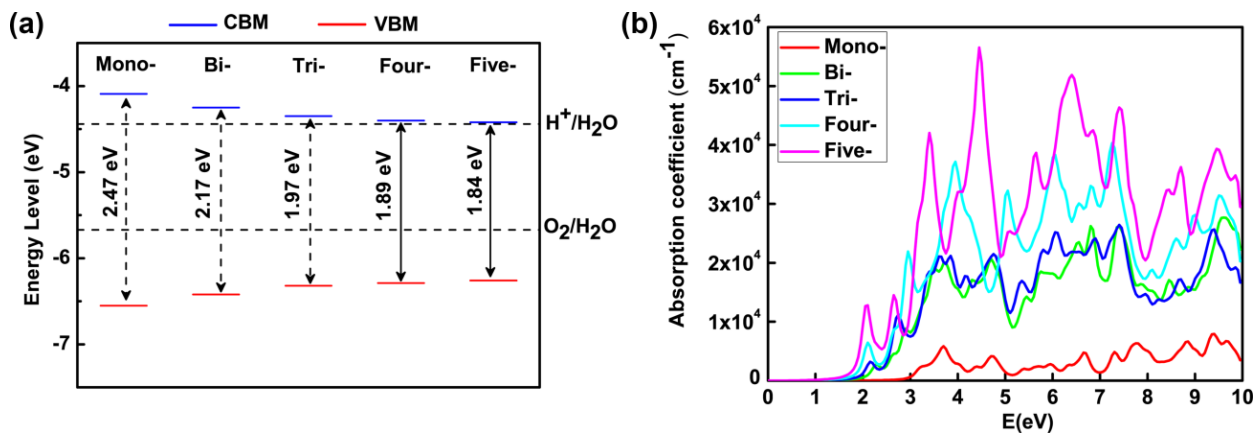

Figure 3. (a) Band edge positions of monolayer and bilayer $C_2N$. The dashed lines are water redox potentials. (b) Optical spectra for few-layer $C_2N$.

## IV. SUMMARY

In conclusion, on the basis of DFT calculations, we have performed a theoretical research on the electronic structure of monolayer and multilayer $C_2N$. We have shown theoretically that few-layer $C_2N$ is a novel category of 2D direct band gap semiconductor and the bandgap of these system can be tuned from 2.47 eV for monolayer to 1.84 eV for five-layer. What is more, based on first principles calculations, we predict few-layer $C_2N$ is a potential metal-free visible-light driven photocatalyst for water splitting. We hope our research to promote the research on layer $C_2N$ and its composite photocatalysts for water spliting.

## AUTHOR INFORMATION

**Corresponding Author**

* E-mail: jlyang@ustc.edu.cn. Phone: +86-551-63606408. Fax: +86-551-63603748 (J. Y.).

**Author Contributions**

The manuscript was written through contributions of all authors. All authors have given approval to the final version of the manuscript.

ACKNOWLEDGMENT

This work is partially supported by the National Key Basic Research Program (2011CB921404), by NSFC (21121003, 91021004, 21233007, 21203099, 21273210), by CAS (XDB01020300), and by USTCSCC, SCCAS, Tianjin, and Shanghai Supercomputer Centers.